\documentclass[12]{article}
\usepackage{graphics,graphicx,amsbsy,amssymb,color}

\newcommand{\UU}{{\boldmath \mbox{$U$}}}

\newcommand{\rr}{{\boldmath \mbox{$r$}}}

\newlength{\defbaselineskip}
\newcommand{\setlinespacing}[1]%
           {\setlength{\baselineskip}{#1 \defbaselineskip}}

\title{\textbf{
Landau Theory of Dynamic Critical Phenomena in  the Rayleigh-B\'{e}nard System}}
\author{Miroslav Grmela \footnote{ e-mail:
miroslav.grmela@polymtl.ca}\\
\'{E}cole Polytechnique  Montr\'{e}al,
  C.P.6079 suc. Centre-ville,\\
 Montr\'{e}al, H3C 3A7,  Qu\'{e}bec, Canada}

 \date{}

\begin{document}

\maketitle

\tableofcontents
\setlength{\parskip}{4mm}

\begin{abstract}

Physics involving more details than hydrodynamics is needed to formulate rate thermodynamics of the Rayleigh-B\'{e}nard system.
The Boussinesq vector field is approached in the space of mesoscopic vector fields similarly as  equilibrium sates are approached  in externally unforced systems in the space of mesoscopic state variables. The approach is driven by gradient of a potential (called a rate entropy). This potential then  provides the rate thermodynamics in the same  way as  the entropy provides thermodynamics for externally unforced systems. By restricting the investigation  to a small neighborhood of the critical point we  can use the rate-thermodynamic version of the Landau theory.

\end{abstract}

\section{Introduction}

Externally unforced macroscopic systems are seen in mesoscopic experimental observations to approach equilibrium states at which no time evolution takes place. In    mathematical models   the approach appears to be driven by a potential called a mesoscopic entropy. In the course of the time evolution the mesoscopic entropy increases and at the equilibrium state reaches its maximum (Maximum Entropy principle - MaxEnt) that   becomes the equilibrium entropy. As a function of  Lagrange multipliers, introduce in order to account for constraints in the maximization, the equilibrium entropy is the fundamental thermodynamic relation on the level of equilibrium thermodynamics.
The route from a \textit{starting mesoscopic dynamical theory (SMT)} to  \textit{equilibrium thermodynamics (ET)}
\begin{equation}\label{eq1}
SMT\,\,\longrightarrow\,\,mesoscopic\,\, entropy\longrightarrow\,\,ET
\end{equation}
has two stages.
The first stage, represented in (\ref{eq1}) by the first arrow,   is passed by investigating solutions inside SMT. In the case of  Boltzmann's investigation SMT is kinetic theory represented by the Boltzmann equation. The Boltzmann mesoscopic entropy arises in the  analysis of its asymptotic solutions (H-theorem). In the case of the GENERIC equation \cite{Gr84}, \cite{Mor}, \cite{GrPhD}, \cite{GO},\cite{OG} governing the time evolution in SMT it is the Hamilton-Jacobi investigation of the intrinsic compatibility of its lift to thermodynamic setting of contact geometry \cite{OGP1},\cite{OGP2},\cite{OGP3}. The second part of the route represented by the second arrow is passed either by following solution to its conclusion or by MaxEnt.

The first arrow in (\ref{eq1}) can also be replaced by various types of arguments that have arisen  in statistical mechanics. In particular, we mention the argument developed by Gibbs \cite{Gibbs} in his formulation of equilibrium statistical mechanics or the argument developed by Jaynes \cite{Jaynes} that is based on viewing the problem of pattern recognition in the phase portrait through the eyes of the information theory. Still another way to find  directly the mesoscopic entropy was introduced by Lev Landau \cite{Landau} in his investigation of critical phenomena. Landau conjectured that in a small vicinity of critical point the mesoscopic entropy is universal. His conjecture was later proven  by Ren\'{e} Thom  \cite{Thom}  and Vladimir Arnold \cite{Arnold}.  MaxEnt analysis of the Landau universal mesoscopic thermodynamic potential  is then the Landau theory of critical phenomena.

Now we turn to externally driven macroscopic systems.
External forces prevent approach to equilibrium states, the level ET in (\ref{eq1}) is inaccessible. The absence of the approach to equilibrium implies the absence of entropy and thus an impossibility to introduce thermodynamics. But even in the presence of external forces there exist  \textit{mesoscopic dynamical theories MT} that are autonomous as the SMT theory, involve the time evolution,  but  take into account less details than SMT and  more details than ET.
For instance the behavior observed in hydrodynamic type observations
of the Rayleigh-B\'{e}nard system (a horizontal layer of a fluid heated from below)  is found to be well described by hydrodynamics represented by Boussinesq equations \cite{Bous}. Hydrodynamics is an autonomous theory that involves less details than, for instance,  the completely microscopic theory (in which fluids are seen as composed of microscopic particles), or possibly other autonomous mesoscopic theories as the  kinetic theories,  that can play the role of SMT. In the case of the Rayleigh-B\'{e}nard system the autonomous mesoscopic dynamical theory MT is hydrodynamics.

Externally driven systems  experience various types of critical phenomena (called dynamic phase transitions and dynamic critical phenomena). This type of behavior is seen in the time evolution inside MT. For example in the Rayleigh-B\'{e}nard system it is the onset of a macroscopic flow in the macroscopically static horizontal layer  of a fluid heated from below. This behavior is seen in the time evolution governed by the Boussinesq equation as an appearance of bifurcations.
It seems that the only way to investigate dynamic critical phenomena in externally driven  systems is to investigate them in the time evolution inside MT. There is however another way.
The fact that MT exists as an autonomous theory implies that the  approach from SMT to MT must exist. The outcome of the first stage of the approach is a potential that then in the second stage drives solutions to fixed points.
The final outcome of the route (\ref{eq1}) is the equilibrium state that is indeed independent of time. In externally driven systems the states approached in the passage SMT $\rightarrow$ MT are (at least in general) time dependent. But what is independent of the time in MT is the vector field generating  the time evolution inside MT (the right hand side of the MT time evolution equation). The information about dynamic critical phenomena that is obtained from an investigation of the time evolution inside MT is in fact  an information obtained from  investigating  the time independent vector field generating it. In order to adapt (\ref{eq1}) to externally driven systems
we reinterpret it as an approach in the space of vector fields
\begin{equation}\label{eq2}
SMT\,\, \longrightarrow rate\,\, mesoscopic\,\, entropy \longrightarrow\,\, MT
\end{equation}
rather than in the state space as in (\ref{eq1}). This reinterpretation,  as well as the terminology that uses the adjective "rate" to distinguish (\ref{eq2}) from (\ref{eq1}), were suggested in \cite{GREE}. The passage (\ref{eq1}) represents thermodynamics, the passage (\ref{eq2}) rate thermodynamics.

Dynamic critical phenomena   can be investigated  in two ways: (i) in the time evolution inside MT (this is the standard way), and (ii) in the context of the rate thermodynamics (\ref{eq2}). Some general relations among  thermodynamics, rate thermodynamics, mesoscopic entropy, rate of mesoscopic entropy, and mesoscopic rate entropy  have been explored in \cite{GREE}). In this paper we adapt the Landau theory of critical phenomena to the rate thermodynamics and illustrate it
 on the example of the Rayleigh-B\'{e}nard system. Similarities and differences between thermodynamics (\ref{eq1}) and rate thermodynamics
(\ref{eq2}) become apparent by comparing the investigation of the Rayleigh-B\'{e}nard system, presented in Section \ref{sec2}, with the investigation of the van der Waals gas recalled in Sections \ref{sec12}, \ref{sec13}.

 Before starting  specific illustrations we make two observations elucidating   the difference between thermodynamics and rate thermodynamics. The first is about  fluctuations. It is well known that the critical behavior that is seen in the time evolution inside MT is also seen in the increase of fluctuations. This type of  manifestation of the criticality is in fact the manifestation displayed in the rate thermodynamics. Indeed, the switch from observing  the behavior in the state space to observing it in the space of vector fields is a switch to  more detailed observations. A need for more details manifests itself in observations in the MT state space in  the increase of fluctuations.

 The second observation elucidating the difference between (\ref{eq1}) and (\ref{eq2})  is about  hierarchy formulations of the time evolution equations in SMT. One of the best known example is the reformulation of the Boltzmann kinetic equation into the Grad hierarchy. In this example the SMT theory is the kinetic theory represented by the Boltzmann kinetic equation. In its Grad reformulation the Boltzmann equation takes the form of the hydrodynamic equations (that play the role of MT) that are however still coupled  to other fields (other moments - in the velocity - of the one particle distribution function) that obey their own time evolution equations. These other fields address details not seen in hydrodynamic fields and they enter the hydrodynamic fields in the vector fields of the hydrodynamic equations. Consequently, they are the fields whose time evolution is followed  in (\ref{eq2}). In the viewpoint of the hierarchical formulation of the time evolution on the SMT level,  the route (\ref{eq2}) is the time evolution whose asymptotic solution is the closure on the MT level of the SMT hierarchy.

\section{Equilibrium and dynamic critical phenomena}\label{sec1}

Citing  Kyozi Kawasaki \cite{Kaw},
\textit{"Critical phenomena occur as a result of a delicate balance of repulsive and attractive interactions among large number of molecules"}. We supplement this   characterization of critical phenomena  by noting that, at least in many macroscopic systems, the attractive interaction are long range and are  in  mathematical formulations generated by  energy while the repulsive interactions are short range and are in  mathematical formulations generated by entropy. The entropy replaces the energy because the complex short range repulsive interactions, involving typically many molecules, cannot be expressed in terms of the chosen mesoscopic state variables.

The particles composing the van der Waals gas   are attracted by  long range weak forces and repulsed by strong but very short range forces. Competition between  these two forces brings about changes on the macroscopic scale in the mass density (gas-liquid phase transitions).

Rayleigh-B\'{e}nard system is a horizontal layer of a fluid heated from below. The flow of the particles composing the fluid is driven by a buoyancy force directed upwards and the friction force due to collisions is directed downwards. Competition between  these two forces brings about  onset of an upward directed motion on the macroscopic scale.

While the van der Waals gas and the Rayleigh-B\'{e}nard fluid have nothing in common, the competition between two microscopic forces that bring about   changes on the macroscopic scale are very similar. We shall follow the route (\ref{eq1}) for the van der Waals gas (in Section \ref{sec12}), the route (\ref{eq2}) for the Rayleigh-B\'{e}nard fluid (in Section \ref{sec13}) and compare them.

\subsection{Mesoscopic theory of the van der Waals gas}\label{sec12}

Van der Waals gas has played and continues to play an important role in both equilibrium and nonequilibrium thermodynamics. Its passing the  route  (\ref{eq1}) has been an important source of inspiration.
We  recall briefly the main contributions in the historical order.

The route (\ref{eq1}) for the van der Waals gas was started by Johannes Diderik  van der Waals \cite{vdW} at its end, i.e. at ET. The state variables on the level of equilibrium thermodynamics are $(E,N,V)$, where
$E$ is the energy, $N$ number of moles, and $V$ the volume. The equilibrium fundamental thermodynamic relation is $S=S(E,N,V)$, where $S$ is the equilibrium entropy. The van der Waals fundamental thermodynamic relation
\begin{equation}\label{vdW1}
S^{(ET)}(E,N,V)=k_BN\frac{5}{2}+Nk_B\ln\left[\left(\frac{E}{N}+A\frac{N}{V}\right)^{\frac{3}{2}}\left(\frac{V}{N}-B\right)\right]
\end{equation}
is a two parameter deformation of the fundamental thermodynamic relation of the ideal gas.   If  $A=0$ and $B=0$ then  (\ref{vdW1}) reduces to  the equilibrium fundamental thermodynamic relation  of the ideal gas. The parameter $A$ represents the influence of the long range attraction and $B$ the excluded volume due to hard core repulsion.

We turn now to a more detailed view of the van der Waals gas.
The second stage on the route (\ref{eq1}) begins with
a mesoscopic entropy  $S^{(SMT)}$, a mesoscopic energy $E^{(SMT)}$, and a mesoscopic number of moles $N^{(SMT)}$ constituting the fundamental thermodynamic relation on the mesoscopic level on which the field $n(\rr)$, denoting the local  number of moles on a mesoscopic scale, plays the role of the state variable.  In the  van Kampen \cite{vanKamp} mesoscopic theory of the van der Waals gas they are the following:
\begin{eqnarray}\label{vdW2}
S^{(SMT)}(n(\rr))&=&-\int_{\Omega}d\rr k_{B}\left( n(\rr)\ln n(\rr) +n(\rr)\ln(1-Bn(\rr))\right)\nonumber \\
E^{(SMT)}(n(\rr))&=& \int_{\Omega} d\rr \left( \frac{3}{2}Tn(\rr) + \int_{\Omega}d\rr' n(\rr)V_{pot}(|\rr-\rr'|)n(\rr')\right)\nonumber \\
N^{(SMT)}(n(\rr))&=&\int_{\Omega} d\rr n(\rr)\nonumber \\
V^{(SMT)}(n(\rr))&=& vol \Omega
\end{eqnarray}
$k_B$ is the Boltzmann constant, $\rr\in \Omega\subset \mathbb{R}^3$ is the position vector, $V_{pot}$ stands for the potential generating  the long range attraction and $B$, the same parameter as in (\ref{vdW1}), is the volume of one particle (excluded volume).
Similarly as in (\ref{vdW1}) the mesoscopic fundamental relation (\ref{vdW2}) is a two parameter deformation of the mesoscopic fundamental thermodynamic relation of the ideal gas. If $B=0$ and
$V_{pot}=0$ then (\ref{vdW2}) becomes the fundamental thermodynamic relation representing the ideal gas on the mesoscopic level on which $n(\rr)$ plays the role of the state variable.

With the potentials (\ref{vdW2}) we introduce the SMT-level thermodynamic potential
\begin{equation}\label{PhivdW}
\Phi^{(SMT)}(n(\rr),\alpha, \beta)=-S^{(SMT)}(n(\rr))+\beta E^{(SMT)}(n(\rr))+ \alpha N^{(SMT)}(n(\rr))
\end{equation}
where $\beta=\frac{1}{T}; \alpha=-\frac{\mu}{T}$, $T$ is the temperature and $\mu$ chemical potential. The MaxEnt passage from (\ref{PhivdW}) to (\ref{vdW1}) is made by two successive Legendre transformations. First, we pass from $\Phi^{(SMT)}(n(\rr),\alpha, \beta)$ to $S^{*(ET)}(\alpha,\beta)$ by the Legendre transformation
\begin{equation}\label{SstarvdW}
S^{*(ET)}(\alpha,\beta)=\Phi^{(SMT)}(\hat{n}(\alpha,\beta,\rr),\alpha, \beta)
\end{equation}
where $\hat{n}(\alpha,\beta,\rr)$ is a solution to $\Phi^{(SMT)}_{n(\rr)}(n(\rr),\alpha, \beta)=0$.  We use the shorthand notation $\Phi_x(x)=\frac{\partial\Phi}{\partial x}$.
From $S^{*(ET)}(\alpha,\beta)$  to
$S^{(ET)}(E,N)$ we pass by the Legendre transformation \\$S^{(ET)}(E,N)=\Phi^{*(ET)}(\hat{\alpha}(E,N),\hat{\beta}(E,N),E,N)$, where
$\Phi^{*(ET)}(\alpha,\beta,E,N)=-S^{*(ET)}(\alpha,\beta)+E\beta +N\alpha$ and $(\hat{\alpha}(E,N),\hat{\beta}(E,N))$ is a solution to
\\$\Phi^{*(ET)}_{\alpha}(\alpha,\beta,E,N)=\Phi^{*(ET)}_{\beta}(\alpha,\beta,E,N)=0$. Finally, the passage from $S^{(ET)}(E,N)$ to $S^{(ET)}(E,N,V)$ is made by requiring that  $S,E,N$ are extensive state variables. This means that $\lambda S^{(ET)}(E,N,V)=S^{(ET)}(\lambda E,\lambda N,\lambda V); \lambda \in\mathbb{R}$. It can be directly verified that  $S^{(ET)}(E,N,V)$ obtained from (\ref{vdW2}) is indeed (\ref{vdW1}).

The physical meaning of the state variable $n(\rr)$ on the SMT level is given by the physical meaning of the potentials (\ref{vdW2}) on the SMT level  and the physical meaning of
\begin{equation}\label{ETlevel}
E=S^{*(ET)}_{\beta}(\alpha,\beta);\,\,N=S^{*(ET)}_{\alpha}(\alpha,\beta)
\end{equation}
on the ET level.

Van Kampen's theory of the van der Waals gas has also  been extended to the first stage on the route  (\ref{eq1}) that involves  the time evolution.
In this paper we do not attempt to introduce SMT theory with the time evolution on the route  (\ref{eq2}) for the Rayleigh-B\'{e}nard fluid and we thus do not need to recall SMT theory with the time evolution in (\ref{eq1}) for the van der Waals fluid. An interested reader can find  an information about mesoscopic dynamical  theories of the van der Waals gas in  \cite{GREE}.

\subsection{Van der Waals critical phenomena}\label{sec13}

Phase transitions and critical phenomena make their appearance in the theoretical formulations presented in the previous sections  as geometrical features of the manifold
\begin{equation}\label{eqmanET}
\mathcal{M}_{eq}^{(ET)}=\{(E,N,V,\Phi^{(ET)}(E,N,V))\in M^{(ET)}\times \mathbb{R}|\Phi^{(ET)}_{E}=\Phi^{(ET)}_{N}=\Phi^{(ET)}_{V}=0\}
\end{equation}
on the level of equilibrium thermodynamics  and  the manifold
\begin{equation}\label{eqmanSMT}
\mathcal{M}_{eq}^{(SMT)}=\{(n(\rr),\Phi^{(SMT)}(n(\rr))\in M^{(SMT)}\times \mathbb{R}|\Phi^{(SMT)}_{n(\rr)}=0\}
\end{equation}
on the mesoscopic SMT level on which the field $n(\rr)$ serves as the state variable.
By $M^{(ET)}$, and $M^{(SMT)}$  we denote the state space used on the level of equilibrium thermodynamics and the SMT level respectively. Similarly, $\Phi^{(ET)}=-S^{(ET)}(E,V,N)+\frac{1}{T}E-\frac{\mu}{T}N$ and $\Phi^{(SMT)}=-S^{(SMT)}(n(\rr))+\frac{1}{T}E^{(SMT)}(n(\rr))-\frac{\mu}{T}N^{(SMT)}(n(\rr))$ is the thermodynamic potential on the level of equilibrium thermodynamics and SMT level respectively.

The geometry of (\ref{eqmanET}) and its thermodynamic interpretation can be found in all textbooks of thermodynamics (see e.g. \cite{Callen}). The manifold (\ref{eqmanET})  offers a  very simple and a very instructive picture of the gas-liquid phase transition.
The picture  is however only  a phenomenological description. For an understanding we have to  look deeper into the microscopic nature of macroscopic systems. The manifold (\ref{eqmanSMT}) is a step in this direction.

The complete geometry of the manifold  (\ref{eqmanSMT}) can only be revealed if the nonlinear partial differential equation $\Phi^{(SMT)}_{n(\rr)}=0$ were completely solved. An incomplete but still very useful information about solutions to $\Phi^{(SMT)}_{n(\rr)}=0$ and consequently about the geometry of the manifold (\ref{eqmanSMT}) can be obtained, following  van Kampen \cite{vanKamp},  by making simplifying assumptions.

If we replace $\rr\in \mathbb{R}^3$ with $r\in\mathbb{R}$  and  if we moreover interpret $r$ as the time $t$, then  equation $\Phi^{(SMT)}_{n(t)}=0$ takes the form of  Newton's mechanical equation  which can be solved (see \cite{vanKamp} and also \cite{PKG}). If in addition we assume that $n=cont.$ then  the thermodynamic potential $\Phi^{(SMT)}$ takes  the form
\begin{equation}\label{PhiL}
\Phi^{(SMT)}(n,\alpha,\beta)=n\ln n+n\frac{d\theta}{dn}-\frac{1}{2}\beta \mathcal{V}_{pot}n^2-\left(\alpha-\frac{3}{2}\ln\frac{\beta}{2\pi}\right)
\end{equation}
where $\mathcal{V}_{pot}=\int d\rr' V_{pot}(|\rr-\rr'|)$ and $\theta(n)=\frac{1-Bn}{B}\left(\ln(1-Bn)-1\right)$.
The critical point
\begin{equation}\label{crpoint}
n^{(cr)}=\frac{1}{3B}; \beta^{(cr)}=\frac{27B}{4\mathcal{V}_{pot}};\alpha^{(cr)}=\frac{1}{2}\ln(3B)+\frac{3}{2}\ln\frac{B}{\mathcal{V}_{pot}}+\frac{3}{4}+4\ln\frac{3}{2}
-\frac{3}{2}\ln(2\pi)
\end{equation}
is a solution of  three equations $\Phi^{(SMT)}_n=0,\Phi^{(SMT)}_{nn}=0,\Phi^{(SMT)}_{nnn}=0$.

Next, we concentrate only  on the behavior in a small neighborhood of the critical point. We begin by making Taylor expansion of
(\ref{PhiL})  at  $n=n^{(cr)}$  (we keep only  terms up to the fourth order)
\begin{equation}\label{LvdW}
\Phi^{(cr)}(\xi,\omega_1,\omega_2,\omega_3)=\omega_1\xi+\frac{1}{2}\omega_2\xi^2+\frac{1}{24}\omega_3\xi^4
\end{equation}
where $\xi=n-n^{(cr)}$,
\begin{equation}\label{omega}
\omega_1= k_1(\alpha-\alpha_{cr})+k_2(\beta-\beta_{cr})  ; \omega_2=k_3(\beta-\beta_{cr})
\end{equation}
The parameters $k_1,k_2,k_3, \omega_3$ are expressed in terms of $\mathcal{V}_{pot}$ and $B$. The critical part
$S^{*(cr)}(\alpha,\beta)=\Phi^{*(cr)}(\hat{\xi}(\alpha,\beta),\alpha,\beta)$
of the entropy (\ref{SstarvdW}) (where $\hat{\xi}(\alpha,\beta)$ is a solution of $\Phi^{(cr)}_{\xi}=0$)
is a generalized homogeneous function
\begin{equation}\label{genhom}
S^{*(cr)}(\alpha,\beta)=\frac{1}{\lambda}S^{*(cr)}(\lambda^{\frac{3}{4}}\alpha,\lambda^{\frac{1}{2}}\beta)
\end{equation}
$\lambda\in\mathbb{R}$.
To prove   (\ref{genhom}) we note that   (\ref{LvdW}) implies \\$\Phi^{(cr)}(\lambda^{-\frac{1}{4}}\xi,\omega_1,\omega_2,\omega_3)=\frac{1}{\lambda}\Phi^{(cr)}(\xi,\omega_1,\omega_2,\omega_3)$.
This means  (since $\omega_3$ remained unchanged) that $S^{*(cr)}(\omega_1,\omega_2)=\frac{1}{\lambda}S^{*(cr)}(\lambda^{\frac{3}{4}}\omega_1,\lambda^{\frac{1}{2}}\omega_2)$.  We get (\ref{genhom}) by noting that $(\omega_1,\omega_2)\leftrightarrows (\alpha,\beta)$ is a linear one-to-one transformation, .

The generalized homogeneity of the critical entropy (\ref{genhom}) together with
(\ref{ETlevel}) then implies the critical behavior seen in equilibrium thermodynamic observations.

Before leaving  van Kampen's theory of the van der Waals gas we emphasize that its
most important and  absolutely essential feature  is the difference in the consideration of the attractive and the repulsive forces. The long range attractive force is a gradient of the energy and the hard core repulsive force is a gradient of the entropy. In other words, the long range attractive force is an energetic force, the short range repulsive force is an entropic force.
Using the terminology introduced in \cite{GREE}, the long range attractive force is a gradient of the  inner energy (the part of the energy that can be expressed in terms of the state variables chosen in the SMT theory  -  the field $n(\rr)$ in the van Kampen's theory) and the short range repulsive forces is a gradient  of the inner energy (the part of the energy that cannot be expressed in terms of the state variable chosen in the SMT theory).  Appearance of  entropic forces in mesoscopic dynamical theories is not unusual. For instance a
well known example of the entropic force  is the rubber-elastic force. The energy involved in very complex excluded-volume type interactions among polymeric chains composing the rubber cannot be expressed in terms of the state variables used in the elasticity theory. The energy generating  such interactions is thus considered in the elasticity theory as an internal energy and is  expressed in terms of entropy.

\subsection{Landau theory of the van der Waals gas critical phenomena}\label{131}

We return to the passage (\ref{eq1}) but we   see it  now through the eyes of Landau \cite{Landau}.  We skip the time evolution in SMT as we already did in Section (\ref{sec12}). We assume that the thermodynamic potential $\Phi^{(SMT)}$ exists  but in this section we also skip its specification. We only assume that it is a potential displaying a critical behavior.  We do not know where the critical point is placed in the state space but we assume that it exists.

The investigation begins with the  critical polynomial (Landau polynomial)
\begin{equation}\label{Lanpol}
\Phi^{cr(SMT)}(\zeta, \varpi_1,\varpi_2,\varpi_3)=\varpi_1\zeta+\varpi_2\zeta^2+\varpi_3\zeta^4
\end{equation}
that is the potential $\Phi^{(SMT)}$  restricted to a small neighborhood of the critical point.
Its universality has been anticipated by  Lev Landau \cite{Landau} and proven rigorously  by
Ren\'{e} Thom \cite{Thom} and Vladimir Arnold \cite{Arnold}.
By $\zeta\in\mathbb{R}$ we denote the order parameter, $\varpi_1\in\mathbb{R}, \varpi_2\in\mathbb{R}, \varpi_3\in\mathbb{R}$  are coefficients of the Landau polynomial. The order parameter is a one dimensional subspace of the state space $M^{(SMT)}$ of the SMT mesoscopic theory in which the critical behavior at the critical point is displayed.
The critical entropy $S^{cr}(\varpi_1,\varpi_2,\varpi_3)$ corresponding to (\ref{Lanpol}) is
\begin{equation}\label{ScrL}
S^{*cr}(\varpi_1,\varpi_2,\varpi_3)=\Phi^{cr(SMT)}(\hat{\zeta}(\varpi_1,\varpi_2,\varpi_3), \varpi_1,\varpi_2,\varpi_3)
\end{equation}
where $\hat{\zeta}(\varpi_1,\varpi_2,\varpi_3)$ is a solution of $\Phi^{cr(SMT)}_{\zeta}=0$.
Still remaining in the setting of the abstract formulation of the Landau polynomial (\ref{Lanpol}) we mention one important property of the critical entropy $S^{*cr}$. Let $\varpi_3$ be fixed, then the the critical entropy $S^{*cr}(\varpi_1,\varpi_2)$ is a generalize homogeneous function
\begin{equation}\label{Lgh}
S^{*(cr)}(\varpi_1,\varpi_2)=\frac{1}{\lambda}S^{*(cr)}(\lambda^{\frac{3}{4}}\varpi_1,\lambda^{\frac{1}{2}}\varpi_2); \,\, \forall \lambda\in\mathbb{R}
\end{equation}
This property of the critical entropy has been proven in (\ref{genhom}). The generalized homogeneity (\ref{Lgh}) implies  the universal critical exponents. It is the main result of the Landau theory.

We shall call the Landau polynomial (\ref{Lanpol}) with a particular choice of $\zeta$, $\varpi_1,\varpi_2,\varpi_3$ a particular realization of the Landau polynomial. The next step in the Landau theory of critical phenomena is  to make a  particular realization of (\ref{Lanpol})  expressing the particular physics of the system under investigation. There are two ways to make it: (i) we find the thermodynamic  potential $\Phi^{(SMT)}(x)$, where $x\in M^{(SMT)}$, identify the critical pint,  and restrict $\Phi^{(SMT)}(x)$ to its small neighborhood, (ii) we use a physical insight to specify $\zeta$, $\varpi_1,\varpi_2,\varpi_3$ directly. We used the first way in Section \ref{sec13} for the van der Waals gas. Now  we use the second way.

In the absence of van Kampen's theory of the van der Waals gas (i.e. in the absence of the thermodynamic potential $\Phi^{(SMT)}(x)$) we could proceed as follows.
We choose  $\zeta=n-n^{(cr)}$, where $n$ is the number of microscopic particles $n\in\mathbb{R}$ composing the fluid, as the order parameter.
In order to find the coefficients $\varpi_1,\varpi_2,\varpi_3$ we recall that
the thermodynamic potential $\Phi^{SMT})$ is a linear combination of the entropy, the energy and the number of moles. On the ET level the energy and the number of moles are state variables and the entropy is expressed in terms of them (see (\ref{vdW1})). In order to obtain the complete thermodynamic potential $\Phi^{(SMT)}$ on the SMT level  we have to express all three potentials $S^{(SMT)}, E^{(SMT)}, N^{(SMT)}$ in terms of the order parameter $\zeta$. The number of moles $N^{(SMT)}$ is linear in $n$ which means that $\varpi_1$ is proportional to $\left(\frac{\mu}{T}-\left(\frac{\mu}{T}\right)^{(cr)}\right)$. In the energy $E^{(SMT)}$ we include only its attractive part that is (due to the involvement of particle-particle interactions)  quadratic in $n$. This means that $\varpi$ is proportional to $\left(\frac{1}{T}-\left(\frac{1}{T}\right)^{(cr)}\right)$. The entropy together with the remaining part of the energy that is included in the entropy  provides $\varpi_3$.

\subsection{Landau theory of the  Rayleigh-B\'{e}nard critical phenomena}\label{sec2}

In the previous two sections we have passed the route (\ref{eq1}) for the van der Waals gas. In this section we come to the main topic of this paper. We begin to pass the route (\ref{eq2}) for the Rayleigh-B\'{e}nard system. Following the example of the van der Waals gas where we started at the end of  the route (\ref{eq1}) we also start at the end of the route (\ref{eq2}).  The theory MT is  the Boussinesq dynamics (e.g.\cite{Ann.Rev.}) in which the state variables are two hydrodynamic fields (local temperature field and the hydrodynamic velocity field) and the time evolution is generated by the  Boussinesq vector field. Its investigation  reveals the onset of a motion on the macroscopic scale in the  initially motionless (on the macroscopic scale) fluid when the imposed temperature gradient reaches critical value $(\nabla T)^{(cr)}$ \cite{Reynolds}.

The next step on the route (\ref{eq2}) is the choice of a more microscopic level SMT.   By analyzing the time evolution in the space of its vector fields we arrive at the rate thermodynamic potential
\begin{equation}\label{PsiRB}
\Psi^{(STM)}(\UU,\nabla T)=-\mathcal{S}^{(SMT)}(\UU)+<\nabla T,\mathcal{E}^{(SMT)}(\UU)>
\end{equation}
generating approach $SMT \longrightarrow  MT$ to the Boussinesq vector field. By $\UU$ we denote the  vector fields on the SMT level, $\mathcal{S}(\UU)$ is the rate entropy, $\mathcal{S}^{(SMT)}(\UU)$ is the rate energy  on the level of  hydrodynamics expressed in terms of $\UU$, and $\nabla T$ is the imposed temperature gradient. If we compare the rate thermodynamic potential (\ref{PsiRB}) with the thermodynamic potential (\ref{PhivdW}) we see that the entropy $S$ is replaced by the rate entropy $\mathcal{S}$, the temperature $T$ by the gradient of the temperature $\nabla T$, and energy $E$ by the rate energy $\mathcal{E}$.
The potential $\Psi^{(SMT)}$ in  (\ref{PsiRB})  provides rate thermodynamics of the Rayleigh-B\'{e}nard fluid as the thermodynamic potential  $\Phi^{(SMT)}$ in (\ref{PhivdW}) provides thermodynamics of the van der Waals fluid. Potentials of the type (\ref{PsiRB}) have already appeared in nonequilibrium thermodynamics in investigations of Rayleigh \cite{Rayleigh} and Onsager \cite{Onsager1}, \cite{Onsager2} (see also \cite{Gyarmati}, \cite{Doi}) where they are called Rayleighians. The relation between the roles that  the potentials (\ref{PsiRB}) play  in rate thermodynamics and in the Rayleigh-Onsager variational principle is discussed in \cite{OGP1}, \cite{OGP2}, \cite{GREE}.

We hope to take the step described above  in a future paper. Here we limit ourselves only to the Landau theory. In the context of the van der Waals gas such limitation  would mean that we start to investigate the critical phenomena of the van der Waals gas  in Section \ref{131}.

If we  see the Rayleigh-B\'{e}nard system only through the eyes of the Boussinesq  equations  (i.e. only inside MT) then there is no potential  displaying  the critical behavior seen in their solutions. Such potential arises when we turn our attention to a dynamical theory  (SMT theory) involving more details than hydrodynamics. Since the Boussinesq theory is an autonomous theory describing well the observed behavior the SMT theory (also an autonomous theory)  has to display the  passage $SMT\longrightarrow MT$. Seeing the approach to MT in  the space of SMT vector fields,  it is the  approach to a fixed point (approach to the Boussinesq vector field).  We assume that this approach is driven by a potential (we call it a rate thermodynamic potential) similarly as the approach to ET in the passage $SMT\longrightarrow ET$ on the route (\ref{eq1}) is driven by the thermodynamic potential. In this paper we see  the rate thermodynamic potential  only through the eyes of the Landau theory.

Our task is to make a  particular realization of the Landau polynomial (\ref{Lanpol}) representing the Rayleigh-B\'{e}nard fluid. The first step is to choose the order parameter $\zeta$. Since we are placing (\ref{Lanpol}) into the context of the route (\ref{eq2}) we know that the order parameter has to be a flux. We choose it to be $\zeta=\nu-\nu^{(cr)}$ where $\nu$ is
 the flux in the vertical direction  on a microscopic scale.  We emphasize that $\nu$ is not the hydrodynamic flux appearing in the Boussinesq  hydrodynamic equations formulated on the MT level. The flux $\nu$ is a flux in the vertical direction on a microscopic scale. It is a flux  of particles composing the fluid. In particular we emphasize that $\nu$ can be different from zero even  if the flux on the hydrodynamic scale is zero. Since the order parameter $\nu$ eventually disappears in the passage to the MT level its exact physical meaning is not needed.  We can regard $\nu$ as a scalar that characterizes the intensity of the microscopic flow.

The next step is to specify the coefficients in the Landau polynomial. We limit ourselves  only to a qualitative specification that is needed in the Landau theory of critical phenomena.
We know that the rate thermodynamic potential is a linear combination of the rate energy  and the rate  entropy. Both these quantities are well known on the level of hydrodynamics (that is  the end of the route (\ref{eq2})). The buoyancy force is expressed in terms of mass density depending on the temperature, the friction force is the Navier-Stokes force expressed in terms of gradients of the velocity field, and the rate entropy is the rate of the local equilibrium entropy expressed in terms of gradients of the hydrodynamic fields. But in the context of the Landau theory we need to express these quantities
as functions of the order parameter $\nu$. The rate of energy is  force multiplied by  flux. Two forces  involved in the Rayleigh-B\'{e}nard fluid are the buoyancy force and the friction force. The latter force involves excluded-volume type interactions  among many particles and its microscopic formulation requires  more details than those that can be expressed in terms of the order parameter $\nu$. Consequently, we include  the friction force in the rate entropy similarly as we included the hard-core type repulsive forces in the van der Waals gas in the entropy. We consider the friction force to be a rate entropic force. Consequently, the friction force  makes its appearance   in the Landau polynomial in the coefficient $\varpi_3$.

Now we turn to
the buoyancy force that originates in the  thermal expansion. On the level of hydrodynamics we express it simply (as it is done in Boussinesq equation) by letting   the mass density  depend  on the temperature. On the SMT level we have only the order parameter $\nu$ as the state variable and we have to therefore express it in terms of $\nu$. We again compare this difficulty with the same type of difficulty we encountered in the analysis of the van der Waals gas. The attractive energy is included  on the ET level  in (\ref{vdW1}) simply by changing the energy $E$ (that is the state variable on the ET level) while its specification in (\ref{vdW2})  on the SMT level  requires a deeper insight into the physics involved. The insight that leads us to express the thermal expansion in terms of $\nu$ is that  the thermal expansion is caused by an increase in  the microscopic flux of the particles composing the fluid. We therefore assume that the buoyancy force is proportional to  $\nu$. This means that the buoyancy force makes its appearance in the coefficient $\varpi_2$.

The first term $\varpi_1 \nu$ in the Landau polynomial (\ref{Lanpol}) is absent since $\nu$ is not constrained. The order parameter $n$ in the Landau theory of the van der Waals fluid is constrained since the total number of microscopic particles $N$ remains constant.

The generalized homogeneity (\ref{Lgh}) implied by the Landau polynomial is thus
\begin{equation}\label{SScrit}
\mathcal{S}^{(cr)}(b)=\frac{1}{\lambda}\mathcal{S}^{(cr)}(\lambda^{\frac{1}{2}}b);\,\,\lambda\in\mathbb{R}
\end{equation}
where $b=\left(\nabla T-(\nabla T)^{(cr)}\right)$. Consequently,
\begin{equation}\label{end}
\nabla \mathfrak{E}\sim \left((\nabla T)^{(cr)}-\nabla T\right)
\end{equation}
where  $\nabla \mathfrak{E}= \mathfrak{E}-\mathfrak{E}^{(cr)}$, $\mathfrak{E}=\frac{\partial \mathcal{S}^{(cr)}}{\partial b}$ is the conjugate of the force $b$ that has the physical interpretation of the  energy flux on the hydrodynamic scale.  The flux  $\mathfrak{E}^{(cr)}=\frac{\partial \mathcal{S}^{(cr)}}{\partial b}|_{b=b^{(cr)}}$ is
 $\mathfrak{E}$ at the critical point.

Finally we mention  a limitation of  the passages  (\ref{eq1}), (\ref{eq2}), and  the Landau theory.  Its root is the observation that    the closer we are to the critical point the more details have to be taken into account in order to formulate an autonomous dynamical theory SMT. Large fluctuations are observed on all levels of description. The starting  dynamical mesoscopic theories  SMT  in (\ref{eq1}), (\ref{eq2})  become inadequate.  One of the consequences of this experimental observation  is that the values of critical exponents obtained by following  (\ref{eq1}), (\ref{eq2})  are only good approximations of the experimentally observed values. The complication brought about by the appearance of large fluctuations    offers however  a possibility to redefine the criticality. Instated of seeing it in  geometrical features of the manifolds (\ref{eqmanET}) and (\ref{eqmanSMT}) (or alternatively by carrying the Gibbs equilibrium statistical mechanics  to the thermodynamic limit $N\rightarrow\infty; V\rightarrow \infty, \frac{N}{V}=const.$ \cite{Ruelle})  one can see  it in the absence of  patterns in the microscopic  phase portraits. This viewpoint of criticality has been introduced in \cite{Kadanoff}, \cite{Kogut} in the context of the Gibbs equilibrium statistical mechanics under the name  renormalization group theory of the critical phenomena. In the context of the mesoscopic  van der Waals theory recalled  above in this section the no-pattern view of criticality is explored  in \cite{GR1}, \cite{GR2}.

\section{Concluding remark}

A theory that has all principal features of the classical thermodynamics can  always be extracted  from  a time  evolution in which gradient of a potential drives trajectories to fixed points.  Mesoscopic  time evolutions of externally unforced systems (e.g. the time evolution governed by the Boltzmann kinetic equation) do have such property, mesoscopic time evolutions of externally driven systems (e.g. the time evolution governed by the Boussinesq equations) do not have it \cite{RevModPhys}. This fact is often interpreted as the  absence of thermodynamics for externally driven systems. A more thorough look into the  microscopic physics involved in externally driven systems reveals however that  the time evolution generating thermodynamics can be found in them in the time evolution of vector fields that generate  the time evolution on more microscopic scales. The fixed point in such time evolution is the vector field generating the mesoscopic  time evolution (e.g. the time evolution governed by the  Boussinesq equations). The possibility of applying  the Landau theory of equilibrium critical phenomena to the dynamic critical phenomena (that arise for instance in the Rayleigh-B\'{e}nard system) is one of the advantages of having thermodynamics of externally driven systems.

In this paper we are making a first step in the introduction of  an alternative (thermodynamic) view of the Rayleigh-B\'{e}nard instability. Instead of viewing it only inside hydrodynamics (MT theory) we see it also inside another theory (SMT theory) that takes into account more details than the MT theory. In this paper we  investigate the appearance of the Rayleigh-B\'{e}nard instability
in the passage $SMT\longrightarrow MT$  only from the point of view of the Landau theory of critical phenomena. We intend to continue this investigation by explicitly formulating the SMT theory and  studying  the approach $SMT\longrightarrow MT$ inside it. Such investigation is expected to provide the complete rate thermodynamic potential $\Phi^{(SMT)}$ driving the approach $SMT\longrightarrow MT$. The investigation of the time evolution inside the SMT theory is also expected    to clarify the relation between the rate thermodynamic view of dynamic critical phenomena that is based on the approach $SMT\longrightarrow MT$  and the conventional view that is based on the analysis of the time evolution inside MT.

\textbf{Acknowledgement}
\\

I would like to thank O\v{g}ul Esen, V\'{a}clav Klika, and  Michal Pavelka,  for stimulating discussions.
\\

\textbf{Conflict of Interest}
\\
The author declares no conflict of interest
\\

\textbf{Data Availability Statement}
\\
No datasets were generated or analysed during the current study


\begin{thebibliography}{}

















\bibitem{Gr84}
Grmela, M., Particle and bracket formulations of kinetic equations Contemp. Math. 28,  125–32 (1984)








\bibitem{Mor}
 Morrison, P.J. A paradigm for joined Hamiltonian and dissipative systems. Physica D,  18, 410–419 (1986)


\bibitem{GrPhD}
 Grmela, M. Bracket formulation of diffusion-convection equations. Physica D,  21, 179–212 (1986)

\bibitem{GO}
Grmela, M., \"{O}ttinger, H.C. Dynamics and thermodynamics of complex fluids: General formulation. Phys. Rev. E, 56, 6620 (1997)


\bibitem{OG}
\"{O}ttinger, H.C.; Grmela, M. Dynamics and thermodynamics of complex fluids: Illustration of the general formalism. Phys. Rev. E
56, 6633 (1997)





\bibitem{OGP1}
O.Esen, M. Grmela, M. Pavelka, On the role of geometry in statistical mechanics and thermodynamics I: Geometrical perspective, J. Math. Phys. 63 (12) (2022)

\bibitem{OGP2}
O. Esen, M. Grmela, M. Pavelka, On the Role of Geometry in Statistical Mechanics and Thermodynamics  II:  Thermodynamic Perspective, J. Math. Phys. 63, (12) 123305 1-21  (2022)


\bibitem{OGP3}
O. Esen, M. Grmela, M. Pavelka, Geometry of Dissipation,  in preparation


\bibitem{Gibbs}
Gibbs, J.W. Collected Works; Longmans Green and Co.: New York, NY, USA, (1984)


\bibitem{Jaynes}
Jaynes, E. T., Foundations of probability theory and statistical mechanics,
in Delaware Seminar in the Foundation of Physics (M. Bunge, ed.). Springer,
New York (1967).

\bibitem{Landau}
D. Landau, On the theory of pase transitions, Zh. Eksp. Teor. Fiz. 7, 19-32, (1937)


\bibitem{Thom}
Thom, R.,  Stabilit\'{e} structurelle et morphog\'{e}n\`{e}se, Paris, Inter \'{E}ditions, (1977)

\bibitem{Arnold}
 V. Arnold, Catastrophe theory, Springer Verlag, (1986)

\bibitem{Bous}
Boussinesq, M.J., Th\'{e}orie de l'intumescence liquide appell\'{e}e onde solitaire ou de translation, se propageant dans un canal rectangulaire. C. R. Acad. Sci. 72: 755–759 (1871)




\bibitem{GREE}
M.Grmela, Thermodynamics and Rate Thermodynamics, J.Stat.Phys. 191:75 (2024)

\bibitem{Kaw}
K. Kawasaki, Slow, intermediate and fast dynamics in condensed matter, Physica A 306, 1-14 (2002)


\bibitem{vdW}
van der Waals, J. D., Thesis (Leiden) (1873)






\bibitem{vanKamp}
van Kampen, N., Condensation of classical gas with long-range attraction, Phys. Rev. 135, A362, (1964)

\bibitem{Callen}
Callen,H.B., Thermodynamics: an introduction to the physical theories of equilibrium thermodynamics and
508 irreversible thermodynamics,Wiley, (1960)


\bibitem{PKG}
Pavelka, M., Klika, V., Grmela, M.,  Multiscale Thermo-Dynamics; De Gruyter: Berlin, Germany,  (2018)


\bibitem{Ann.Rev.}
Bodenschatz, E., Pesch, W., Recent developments in Rayleigh-B\'{e}nard convection, Ann. Rev. Fluid Mech. 32, 709-778 (2000)


\bibitem{Reynolds}
 Rayleigh, Lord, On the convective currents in a horizontal layer of fluid when the higher temperature is on the under side,  Philosophical Magazine. 6th series. 32 (192): 529–546 (1916)



\bibitem{Rayleigh}
Rayleigh, Lord. Proc. Math. Soc. London, 4, 357 (1873)





\bibitem{Onsager1}
L. Onsager. Reciprocal relations in irreversible processes I, II. Physical
Review, 37(4),405, 38(12):2265, (1931)

\bibitem{Onsager2}
 L.Onsager, S. Machlup, Fluctuations and Irreversible Processes Physical
Review. 91 (6), 1505–1512 (1953)

\bibitem{Gyarmati}
 Gyarmati, I. Non-Equilibrium Thermodynamics; Springer: Berlin, Germany,
(1970).


\bibitem{Doi}
 M.Doi, Onsager’s variational principle in soft matter, J.Phys. Condensed
matter, 23, 284118 (2011)



\bibitem{Ruelle}
Ruelle, D. Thermodynamic Formalism, 2nd ed.; Cambridge University Press: Cambridge, UK, (2010)

\bibitem{Kadanoff}
 Kadanoff, L. P.,  Scaling laws for Ising models near $T_c$, Physics Physique Fizika. 2 (6): 263 (1966)




\bibitem{Kogut}
 K.G.Wilson, Phys. Rev. B, 4, 3174, 3184, (1971)


\bibitem{GR1}
M. Grmela, Renormalization of the Van derWaals theory of critical phenomena, Phys. Rev. A, 14, 1781-1789
510 (1976)


\bibitem{GR2}
 M. Grmela, V. Klika, M. Pavelka, Dynamic and Renormalization-Group Extensions of the Landau Theory of Critical Phenomena, Entropy, 22, 978, (2020)


\bibitem{RevModPhys}
Cross, M.C., Hohenberg, P.C., Pattern formation out of equilibrium, Rev. Mod. Phys., 65, 851-1112 (1993)




\end{thebibliography}
\end{document}